\documentclass[prb,showpacs,twocolumn,amsmath]{revtex4}
\usepackage{latexsym}

\newcommand{\be}{\begin{equation}}

\newcommand{\ee}{\end{equation}}

\newcommand{\giro}[1]{\stackrel{\circ}{#1}}

\begin{document}

\title{ Scissors Modes:The first overtone}




\author{Keisuke Hatada$^{1,2}$, Kuniko Hayakawa$^{2,3}$, Fabrizio Palumbo$^{2}$}

\affiliation{$^1$ Instituto de Ciencia de Materiales de Arag\'on,
CSIC-Universidad de Zaragoza, 50009 Zaragoza, Spain,\\
$^2$INFN Laboratori Nazionali di Frascati, c.p. 13, I-00044 Frascati, Italy,\\
$^3$Centro Fermi, Compendio Viminale, Roma I-00184, Italy}


{\small e-mail: {\tt hatada@unizar.es,hayakawa@lnf.infn.it,palumbof@lnf.infn.it}}


\date{\today}

\begin{abstract}
Scissors modes were predicted in the framework of the Two-Rotor Model. This model has an intrinsic harmonic spectrum, so that the level above the Scissors Mode, the first overtone,  has excitation energy twice that of the Scissors Mode. Since the latter is of  the order of 3 MeV in the rare earth region, the energy of the overtone is below threshold for nucleon emission, and its width should remain small enough for the overtone  to be observable. We find that  
$B(E2)\uparrow_{overtone} = {3 \over 64 \, \theta_0^{2}}B(E2)\uparrow_{scissors}$, where $\theta_0$ is the zero-point oscillation amplitude, which in the rare earth region is of order $ 10^{-1}$.
\end{abstract}

\maketitle

The Scissors Mode has  an excitation energy of the order of $ 3 \, MeV$ in the rare earth region~[\onlinecite{Bohl,Ende}]. The model which led to its prediction, the Two-Rotor Model~[\onlinecite{LoIu}],  has  the spectrum of a planar harmonic oscillator, with some constraints on the states to be discussed below.  The first overtone should then have an energy of the order of $ 6 \, MeV$, below threshold for nucleon emission. As a consequence its  width  should  remain of purely electromagnetic nature, and   be small enough for this mode to be observable, even though Scissors Modes  have a modest collectivity and are substantially fragmented~[\onlinecite{Bohl, Ende}]. 

The possible occurrence of the first overtone has been considered in Ref.[\onlinecite{Sun}], but to our knowledge has not been thoroughly investigated so far.  The main reason  is perhaps that  its excitation amplitude is expected to be  too small. Indeed excitation amplitudes in the Two-Rotor Model    are proportional to powers of $\theta_0$, the amplitude of the zero point oscillation,  which in the rare earth region is of order $10^{-1}$. 
Now $ B(M1)\uparrow_{scissors}  \, \sim 1/ \theta_0^2$, but $B(E2)\uparrow_{scissors} \, \sim \theta_0^2$, and the first overtone needs to be excited by an $E2$ multipole. All other methods used in the study of Scissors Modes,  the schematic  Random Phase Approximation~[\onlinecite{Suzu}], the Interacting Boson model~[\onlinecite{Iach}], the sum rule method~[\onlinecite{Lipp}] and a geometrical model~[\onlinecite{Roho}] give similar results.  

The proportionality  $B(M1)\uparrow_{scissors}\sim B(E2)\uparrow_{scissors}$  has been observed  experimentally~[\onlinecite{Rang}] (even though  $B(M1)\uparrow_{scissors}$  in reality scales as $\delta^2$  instead of  $\delta^{{3 \over 2}}$ as predicted by 
the Two-Rotor Model if the dependence of the moment of inertia on the deformation parameter of Ref.[\onlinecite{LoIu2}] is adopted).   This  is regarded as a significant evidence of the Scissors nature of the low lying magnetic transitions and their collectivity. We expect that such a proportionality should hold for the $E2$ strength of excitation of the first overtone as well. Even if this strength were small, its study might contribute to a deeper assessment of  the nature and collectivity of the Scissors Modes.  We then decided to start an investigation  of the overtone in the Two-Rotor  Model, because this model allows a first insight into the relevant dynamics in a simple framework. We thus came to the surprising result  that  $B(E2)\uparrow_{overtone}$ is of  zero order in the expansion with respect to $\theta_0$, and precisely
 \be
 B(E2)\uparrow_{overtone} = {3 \over   64 \, \theta_0^2} \, B(E2)\uparrow_{scissors}\,. \label{BE2}
 \ee
 In view of the smallness of $\theta_0$, this amplitude is quite substantial. The above result might be relevant to  some of the other electrically charged 
   systems for which Scissors Modes have been predicted:   metal clusters~[\onlinecite{Lipp1}], quantum dots~[\onlinecite{Serr}]
and, in particular  crystals~[\onlinecite{Hata}], for which an expansion in powers of $\theta_0$ holds.
In all these systems one of the blades of the scissors must be identified with a moving cloud of particles (electrons in metal clusters and
quantum dots,   an atom in  a cell in crystals) and the other one with a structure at rest, the lattice. 

In order to make the paper self contained we report the main features  of the Two-Rotor Model. Its classical hamiltonian is 
\be
H= \frac{1}{2 \, {\mathcal I}_n} {\vec I}_n^2 +  \frac{1}{2\, {\mathcal I}_p} {\vec I}_p^2 +V \label{classical}
\ee
where $ {\vec I}_n,  {\vec I}_p, {\mathcal I}_n, {\mathcal I}_p$ are the angular momenta and the moments of inertia of the neutron and proton bodies  assumed to have ellipsoidal shape and $V$ their interaction potential.  Introducing the total angular momentum ${\vec I}$ and the vector ${\vec S}$
\be
{\vec I} = {\vec I}_n + {\vec I}_p\,, \,\,\, {\vec S} = {\vec I}_n -  {\vec I}_p
\ee
 the hamiltonian (\ref{classical}) can be rewritten as the sum of the rotational hamiltonian of the nucleus as a whole plus the hamiltonian of the intrinsic motion
  \be
 H={ {\vec I}^2 \over 2  {\mathcal I} } +H_{intr} 
 \ee
 where 
 \be
 {\mathcal I} = { 4 \, {\mathcal I}_p{\mathcal I}_n\over {\mathcal I}_p+{\mathcal I}_n}
 \ee
\be
H_{intr}=  \frac{ 1 } {2 \, {\mathcal I}}\, {\vec S}^2 + 
 \frac{ {\mathcal I}_n - {\mathcal I}_p } {4 \, {\mathcal I}_n {\mathcal I}_p } 
\, {\vec I}\cdot {\vec S} +V \,.
\ee
We assume the potential to depend only on the angle $2\theta$ between the symmetry axes ${\hat \zeta}_n, {\hat  \zeta}_p$ of the rotors
\be
\cos (2\theta) = {\hat \zeta}_n \cdot {\hat  \zeta}_p\,.
\ee
It is therefore convenient to introduce this variable together with a set of other variables which identify the axes ${\hat \zeta}_n , {\hat  \zeta}_p$.
We chose the Euler angles $ \alpha, \beta, \gamma$ of the intrinsic frame defined by
\be
{\hat \xi}= \frac{{\hat \zeta}_n \times {\hat  \zeta}_p}{ \sin (2 \theta)}   \,, \,\,\, {\hat \eta} = \frac{{\hat \zeta}_n - {\hat  \zeta}_p}{2 \sin \theta}   \,, \,\,\,  {\hat \zeta}= \frac{{\hat \zeta}_n +{\hat  \zeta}_p}{2 \cos \theta}\,.
\ee
The correspondence $  \left( {\hat \zeta}_n, {\hat  \zeta}_p \right)  \leftrightarrow 
 \left( \alpha, \beta, \gamma, \theta \right) $ is one-to-one and regular for $ 0< \theta < {\pi}/{2}$. These variables are not sufficient to describe all the configurations of the classical system, but describe uniquely the quantum system owing to the constraint
\be
{\vec I}_n \cdot {\hat \zeta}_n = {\vec I}_p \cdot {\hat \zeta}_p=0
\ee
appropriate to quantum bodies with axial symmetry. These constraints are automatically satisfied if the wave functions depend on ${\hat \zeta}_n ,{\hat \zeta}_p $ 
only. Quantization can be obtained assuming the standard representation  for the total angular momentum ${\vec I}$, and  for ${\vec S}$ the realization
\be
S_{\xi}= i \frac{\partial}{\partial \theta}\,, \,\,\,  S_{\eta}= - \cot \theta I_{\zeta} \,, \,\,\, S_{\zeta}= - \sin \theta I_{\eta}
\ee
where $I_{\xi}, I_{\eta}$ and $I_{\zeta}$ are the components of the total angular momentum on the intrinsic axes.

The Two-Rotor Model was reformulated~[\onlinecite{DeFr}] adopting a more appropriate quantization procedure,  and  including the realistic case in which the neutron rotor is bigger than that of protons. The resulting intrinsic hamiltonian is
 \begin{eqnarray}
H_{intr}&=& { 1\over 2 {\mathcal I}}
 \Bigg[ - { d^2 \over d \theta^2 } - 2 \cot ( 2\theta) {d \over d \theta}+ \cot^2 \theta \, I_{\zeta}^2
\nonumber\\
 & +&  \tan^2 \theta  I_{\eta}^2  \Bigg]  +{{\mathcal I}_n - {\mathcal I}_p \over  4 \, {\mathcal I}_p {\mathcal I}_n}\Bigg[ -\tan \theta \, I_{\zeta} I_{\eta} - \cot \theta  \, I_{\eta} I_{\zeta}
\nonumber\\
&+ &
i I_{\xi} { d \over d \theta}\Bigg]  + V(\theta)\,. 
\end{eqnarray}
We separate the range of $\theta$ in two regions
\be
s_I= s(\theta) s\left({\pi \over 4}-\theta\right), \,\, s_{II}= s\left({\pi \over 2}-\theta \right) \,s\left(\theta -  {\pi \over 4} \right), 
\ee
where $s(x)$ is the step function: $ s(x)=1, x>0$ and zero otherwise. They are obtained from each other by  the reflection of $\theta$ with respect to ${\pi / 4}$. It is convenient to introduce the notation
\be
R_{\theta}f(\theta) = \giro{f}(\theta)
\ee
where
\be
\giro{f}(\theta) = f \left( {\pi \over 2} - \theta \right)\,,
\ee
so that $ \giro{s}_I = s_{II} $.
 We assume $\giro{V}=V$, as appropriate to the geometry of the system.
 Since we know that the angle between the neutron-proton axes is 
 very small  we can assume for the potential a quadratic approximation
\be
V= {1\over 2} C \, \theta_0^2 \, x^2 s_I + {1\over 2} C \,\theta_0^2 \, y^2 s_{II}
\ee
where
\be
\theta_0 = ({\mathcal I}C)^{-{ 1\over 4}}\,, \,\,\,x= {\theta \over \theta_0}\,,  \,\,\, y= { {\pi \over 2} - \theta \over \theta_0}\,.
\ee
The intrinsic hamiltonian is then invariant with respect to the transformation
\be
R= R_{\xi} \left({\pi \over 2} \right) R_{\theta} 
\ee
where $R_{\xi} $ is the rotation operator around the $\xi$-axis,
so that we can study the eigenvalue equation separately in the regions $I,II$.
 The linear derivative in the first term can be eliminated by  the transformation
\be
( U \Phi )(\theta)= { 1 \over  \sqrt{2 \sin(2 \theta)}} \,  \Phi'(\theta)\,. \label{transformation}
\ee
Neglecting  terms of order $\theta_0^{-1}$ we get
 \begin{eqnarray}
H_{intr}' &=& U H_{intr} U^{-1}=  { 1 \over 2  {\mathcal I} } \Bigg[ - { d^2 \over d \theta^2 } -\left( 2+  \cot^2 ( 2\theta)\right)
\nonumber\\
&+& \cot^2 \theta \, I_{\zeta}^2 + \tan^2 \theta  I_{\eta}^2  \Bigg] +V(\theta)\,.
\end{eqnarray}
We then write accordingly
\be
H_{intr}' \approx H_I s_I + H_{II} s_{II}
\ee
where, setting $\hbar=1$ 
\begin{eqnarray}
H_I&=&{1\over 2} \omega \left[ - {d^2 \over d x^2} + { 1\over x^2} \left( I_{\zeta}^2 - { 1\over 4} \right) + x^2 \right]
\nonumber\\
H_{II} &= &{1\over 2} \omega \left[ - {d^2 \over d y^2} + { 1\over y^2} \left( I_{\eta}^2 - { 1\over 4} \right) + y^2 \right]
\end{eqnarray}
with
\be
\omega = \sqrt{C \over {\mathcal I}}\,.
\ee
The eigenfunctions and eigenvalues of $H_I$ are~[\onlinecite{DeFr}]
\begin{eqnarray}
\varphi_{Kn}(x) &=& \sqrt{ { n! \over (n+K)! \, \theta_0}} \, x^{K+{1 \over2}} \, L_n^K\left(x^2 \right) e^{-{ 1\over 2} x^2}
\\
\epsilon_{nK}&=& \omega (2n +K +1)\,.
\end{eqnarray}
 $ L_n^K $ are Laguerre polynomials and the wave functions $\varphi$ are
normalized according to
\be
\int_0^{\infty} dx \, \left(\varphi_{Kn}(x)\right)^2 = { 1\over 2}\,.
\ee
Even if the nucleus in its ground state has axial symmetry, this symmetry is in general lost in excited states, so that the component of angular momentum along any intrinsic axis is not conserved, resulting in a superposition of intrinsic states with different $K$-quantum number. The only states which have been theoretically analyzed  so far are not affected by $K$-mixing. They are the ground state, $I=K=n=0$, the Scissors Modes, $I=1,2, K=1, n=0$, and the state $I=K=0, n=1$, which cannot be excited  by electromagnetic radiation.

 The total wavefunctions must respect the $r$-symmetry with respect to both the neutron and proton axes:
configurations of the nucleus differing by independent  rotations through $\pi$ around the $\xi$- axis of the ${\vec{\zeta}}_n, {\vec{\zeta}}_p$ vectors are indistinguishable. Enforcing this symmetry  one finds two conditions~[\onlinecite{DeFr}]. The first has been solved in general, and requires that  the wave functions  have the form
\be
\Psi_{IM \sigma} = \sum_{K \ge 0}{\mathcal F}^J_{MK}(\alpha, \beta, \gamma) \Phi_{IK\sigma}(\theta) \label{Intr}
\ee
where
\be
{\mathcal F}^I_{MK}= \sqrt{{2J+1}\over 8 ( 1 +\delta_{K0}) \pi^2 } \left( {\mathcal D}^J_{MK} +(-1)^J {\mathcal D}^J_{M-K}   \right) \,.
\ee
$I,M$ are the nucleus angular momentum and its component on the $z$-axis of the laboratory system, and $\sigma$ labels the states. 
We impose the normalization
\be
\int_0^{2\pi} d\alpha \int_0^{\pi}d \beta \int_0^{2 \pi }d\gamma \int_0^{{\pi \over 2}}d \theta \, |\Psi_{IM\sigma}|^2 =1\,.
\ee
Notice that the normalization of the $\Phi$ in Eq.(\ref{Intr}) is different from that in Ref.~[\onlinecite{DeFr}]. 
The second condition coming from $r$-invariance constrains the parity of the intrinsic functions with respect to $R_{\theta}$ and must be worked out state by state.

 The ground state and the Scissors Mode are  labelled by  $ \sigma=0,1$ respectively, and their intrinsic  wave functions are 
  \begin{eqnarray}
\Phi_{000}&=& \varphi_{00} \, s_I +\giro {\varphi}_{00}  s_{II}
\nonumber\\
\Phi_{111}&=&  \varphi_{10} \, s_I - \giro{\varphi}_{10}  s_{II}\,.
\end{eqnarray}
 Let us now study the  first overtone, labelled by $\sigma=2$.  The hamiltonian $H_{II}$ couples states with $K=0,2$.  It is then easy to see that the total eigenfunction must involve  a superposition of $\varphi_{01}$ and $\varphi_{20}$. These states are degenerate and decoupled in region $I$. In order to find which superposition is an eigenfunction of the total hamiltonian  we diagonalize $I_{\eta}^2$
\begin{eqnarray}
 I_{\eta}^2 \, {\mathcal G}_{M0} &=& 0\,,   \, \,\,\, {\mathcal G}_{M0}= {1\over 2} \left({\mathcal F}_{M0}^2 + \sqrt 3 \, {\mathcal F}_{M2}^2  \right) \,, 
\nonumber\\
  I_{\eta}^2 \, {\mathcal G}_{M2}&= &4 \,{\mathcal G}_{M2}\,, \,\,\,
{\mathcal G}_{M2} = {1\over 2} \left( \sqrt 3 \,{\mathcal F}_{M0}^2 - {\mathcal F}_{M2}^2  \right).
\end{eqnarray}
 The total eigenfunction  must then be an appropriate  superposition of 
 ${\mathcal G}_{M0} \giro{\varphi}_{01}, {\mathcal G}_{M2} \giro{\varphi}_{20}$
  in region $II$ and  of 
  ${\mathcal F}_{M0}^2 \varphi_{01}, {\mathcal F}_{M2}^2 \varphi_{20}$
   in region $I$. It is easy to verify that the intrinsic wave functions  which satisfy the constraints~[\onlinecite{DeFr}] of $r$-invariance are
\begin{eqnarray}
\Phi_{202}&=&{ 1\over \sqrt 2} \left[ \varphi_{01} \, s_I - { 1\over 2} \left( \sqrt{3} \giro{\varphi}_{20} + \giro{\varphi}_{01} \right) s_{II}  \right] 
\nonumber\\
\Phi_{222}&=& { 1\over \sqrt 2}\left[   \varphi_{20} \, s_I + {1\over 2}  \left(  \giro{\varphi}_{20} - \sqrt{ 3} \giro{\varphi}_{01} \right) s_{II}  \right] \,.
\end{eqnarray}
The different normalization of the $\Phi$ should be kept in mind in a comparison with Ref.~[\onlinecite{DeFr}].

The collective motion of the first overtone has a simple geometrical description in the intrinsic frame: in region $I$  it is a superposition of the state $\varphi_{01}$, which is a kind of breathing mode, and of  the state $\varphi_{20}$, which is a relative rotation of the neutron-proton axes as in the Scissors Mode but with angular momentum $K=2$.
We already mentioned that  the  spectrum of the  Two-Rotor Model is identical to that of the planar harmonic oscillator. We  remark however that the first and second excited states of the planar harmonic oscillator have degeneracy  2 and 3 respectively, while all the intrinsic states of the Two-Rotor Model discussed so far are non degenerate because of the $r$-symmetry.

The surprising feature of our result  is that the  $E2$ excitation amplitude of the first overtone gets a nonvanishing contribution  to zero order in the expansion with respect to $\theta_0$. Indeed 
the quadrupole operator  to this approximation is~[\onlinecite{DeFr}]
\begin{eqnarray}
M(E2,\mu) &=& e \, Q_{20}  \left[  {\mathcal D}^2_{\mu 0}  \left(s_I - { 1\over 2} \, s_{II}  \right) \right.
\nonumber\\
&&\left. +  {1\over 2} \sqrt {3 \over 2} \left( {\mathcal D}^2_{\mu 2} + {\mathcal D}^2_{\mu -2}   \right) s_{II} \right]
\end{eqnarray}
where $e \, Q_{20}$ is quadrupole moment in the intrinsic frame.
We immediately see that to zero order in $\theta_0$ we  cannot excite the Scissors Modes from the ground state,  but we can excite the first overtone, with the amplitude
\begin{eqnarray*}
&&\langle \Psi_{2M2}| M(E2, \mu)| \Psi_{000}\rangle= { 1 \over 4 \sqrt2} \, e \, Q_{20} \langle \varphi_{20}|\varphi_{00} \rangle
\\
&&
\times \langle  -\sqrt3 \, {\mathcal F}_{M_20}^2 +  {\mathcal F}_{M_22}^2 |\sqrt{3\over 2}\left(  {\mathcal D}^2_{\mu 2} + {\mathcal D}^2_{\mu -2} \right)
-   {\mathcal D}^2_{\mu0} | {\mathcal F}_{00}^0\rangle\,.
\end{eqnarray*}
 Notice that this amplitude  is entirely due to the $K=2$ component of the wave function, because $ \langle \varphi_{01}|\varphi_{00} \rangle=0  $. 
 Finally
\be
\langle \Psi_{2M2}| M(E2, \mu)| \Psi_{000}\rangle= { 1 \over 16} \sqrt{3 \over 10} \, e \, Q_{20} \, C_{002\mu}^{2M}
\ee
where $C^{2M}_{002\mu}$ is a Clebsch-Gordan coefficient.We thus get Eq.(\ref{BE2}).

While the first overtone cannot be excited by a $M1$ transition, it can decay to the Scissors Mode through such a transition to order $\theta_0^{-2}$. The  $E2$ decay to the Scissors Mode can instead only occur to order $\theta_0^2$ and will be neglected.
 After the  excitation of the first overtone one should then observe photons of energy $2\omega$ and $\omega$ as well. Let us then 
  evaluate the $M(M1)$ transition amplitude.
The  magnetic dipole moment in the intrinsic frame is
\be
 M(M1, \nu) = { e\over 2m} {\sqrt  { 3 \over 4 \pi}} S_{\nu}\,.
\ee
Then in the laboratory frame we have
\begin{eqnarray} 
 M(M1, \nu) &=&- { 1 \over \sqrt 2}\left( {\mathcal D}^1_{\nu 1}- {\mathcal D}^1_{\nu -1}  \right) {\mathcal M}
 \nonumber\\
 &\times &
\left( s_I - s_{II} \right) { d \over d \theta}
\end{eqnarray}
where
\be
{\mathcal M} =  i \sqrt {3 \over 16 \pi}  \left( g_p - g_n \right) { e \over 2m_p} 
\ee
$g_n,g_p$ being the orbital gyromagnetic factors of neutrons, protons respectively and $m_p$ is the proton mass. 
We need an approximate expression of the derivative operator in regions $I$ and $II$
\begin{eqnarray}
\theta_0 \, U{d \over d \theta}U^{-1} &\approx &{d \over d x} - { 1\over 2 x}\,, \,\,\, \mbox{in region I}
\nonumber\\
\theta_0 \, U{d \over d \theta}U^{-1} &\approx &- {d \over d y} + { 1\over 2 y} \,, \,\,\, \mbox{in region II}\,.
\end{eqnarray}
Setting
\be
\nabla_{\theta} ={d \over d \theta} - { 1 \over 2 \theta} 
\ee
by a straightforward calculation we get
\be
\langle \Psi_{IM_2 2}|M(M1,\nu) | \Psi_{1M_1 1}\rangle = - { \pi \over \sqrt 3} \, {\mathcal M} \, {\mathcal T}
\ee
where
\begin{eqnarray}
{\mathcal T}&=& \langle {\mathcal F^2_{M_20}}|  {\mathcal F}^1_{\nu1} |{\mathcal F}^1_{M_11} \rangle
 \langle \varphi_{01}- \sqrt 3 \, \varphi_{20}|\nabla_{\theta} |   \varphi_{01} \rangle \nonumber\\
&+&
 \langle {\mathcal F^2_{M_22}}|  {\mathcal F}^1_{\nu1} |{\mathcal F}^1_{M_11} \rangle
\sqrt 3 \, \langle \varphi_{20} -  \varphi_{01}| \nabla_{\theta}| \varphi_{01} \rangle\,.
\end{eqnarray}
At variance with the $E2$ excitation of the first overtone both intrinsic components $K=0, n=1$ and $K=2, n=0$ contribute to this decay amplitude.
We thus find
\be
\langle \Psi_{IM_2 2}|M(M1,\nu) | \Psi_{1M_1 1}\rangle  = {\sqrt 2 + \sqrt 3 \over 2 \sqrt{ 10} \, \theta_0} \, C^{2 M_2}_{1 M_1 1 \nu}  \, {\mathcal M},
\ee
and  we can finally relate the $M1$ decay strength of the overtone to the Scissors mode with the $M1$ strength of the Scissors Mode  excitation
\be
B(M1; overtone \rightarrow scissors) \approx { 1\over 4} B(M1)\uparrow_{scissors}\,.
\ee
 A distinctive feature of the overtone in the Two-Rotor Model is the mixing of intrinsic  states with different $K$-quantum number, which is necessary to respect the $r$-symmetry. This mixing is determined by the different form that the intrinsic hamiltonian takes in regions $I$ and $II$. It would be very interesting to investigate if   the  approches~[\onlinecite{Suzu}],[\onlinecite{Iach}],[\onlinecite{Lipp}],[\onlinecite{Roho}],   will confirm the structure of the overtone we found in the Two-Rotor Model.  Indeed so far microscopic calculations, while reproducing some experimental features as fragmentation, which are outside the possibility of a semiclassical model, agree with the Two-Rotor Model about  the nature of the Scissors Modes.  In particular  the Interacting Boson Model, in the semiclassical approximation obtained using coherent states,   exactly reproduces~[\onlinecite{Diep}] the hamiltonian of the Two-Rotor Model in region $I$. We would be very surprised if  the agreement would not  extend  to region $II$, and we think it would  be very interesting to know the exact results for   the overtone in the Interacting Boson Model.


 \end{document}